\catcode`\@=11
\def\citen#1{\if@filesw \immediate\write \@auxout {\string\citation{#1}}\fi%
\@tempcntb\m@ne \let\@h@ld\relax \def\@citea{}%
\@for \@citeb:=#1\do {\@ifundefined {b@\@citeb}%
    {\@h@ld\@citea\@tempcntb\m@ne{\bf ?}%
    \@warning {Citation `\@citeb ' on page \thepage \space undefined}}%
    {\@tempcnta\@tempcntb \advance\@tempcnta\@ne
    \setbox\z@\hbox\bgroup\ifcat0\csname b@\@citeb \endcsname \relax
    \egroup \@tempcntb\number\csname b@\@citeb \endcsname \relax
    \else \egroup \@tempcntb\m@ne \fi \ifnum\@tempcnta=\@tempcntb
    \ifx\@h@ld\relax \edef \@h@ld{\@citea\csname b@\@citeb\endcsname}%
    \else \edef\@h@ld{\hbox{--}\penalty\@highpenalty
    \csname b@\@citeb\endcsname}\fi
    \else \@h@ld\@citea\csname b@\@citeb \endcsname \let\@h@ld\relax \fi}%
\def\@citea{,\penalty\@highpenalty\hskip.13em plus.13em minus.13em}}\@h@ld}
\def\@citex[#1]#2{\@cite{\citen{#2}}{#1}}%
\def\@cite#1#2{\leavevmode\unskip\ifnum\lastpenalty=\z@\penalty\@highpenalty\fi%
  \ [{\multiply\@highpenalty 3 #1%
  \if@tempswa,\penalty\@highpenalty\ #2\fi}]}   %
\makeatother 

\catcode`\@=12
\newcommand\hepv[1]{#1}

\def\A             {{\rm A}}
\def\aff           {affine Lie algebra}
\def\alg           {algebra}
\def\apc           {{{\wfont A}_{\vec p,\tC}}}
\def\apC           {{{\mathfrak A}_{\vec p,\tC}}}

\def\auto          {automorphism}
\def\bearll        {\begin{array}{ll}}
\def\bc            {boundary condition}
\def\Bc            {Boundary condition}
\def\be            {\begin{equation}}

\def\calb          {{\cal B}}
\def\calc          {{\cal C}}
\def\calg          {{\cal G}}
\def\calh          {{\cal H}}

\def\calhl         {{\cal H}_\lambda}
\def\calhm         {{\cal H}_\mu}
\def\calhmv        {\vec{\cal H}_\muv}
\def\calhv         {{\cal H}_\vac}
\def\call          {{\rm L}}
\def\calm          {{\cal M}}
\def\cals          {{\cal S}}
\def\calU          {{\cal U}}

\def\cb            {chiral block}
\def\Cb            {Chiral block}

\def\cft           {conformal field theory}
\def\Cft           {CFT}

\def\cfts          {conformal field theories}
\def\Cfts          {CFTs}
\def\chii          {\raisebox{.15em}{$\chi$}}
\def\chir          {\mbox{$\wfont A$}}
\def\chiR          {{\wfont A}}

\def\chirb         {\mbox{$\bar{\wfont A}$}}
\def\chirB         {\bar{\wfont A}}

\def\cla           {classifying algebra}
\def\clA           {\mbox{$\calc(\chiR)$}}
\def\clAb          {\mbox{$\calc(\chirB)$}}

\def\class         {classification}
\def\complex       {{\dl C}}
\def\Con           {Conformal }
\def\corfu         {correlation function}
\def\Corfu         {Correlation function}
\def\CR            {{\cal R}}
\def\cvo           {chiral vertex operator}
\def\dim           {dimension}

\def\dl            {\mathbb }
\def\dyd           {Dynkin diagram}
\def\ee            {\end{equation}}
\def\eE            {{\rm e}}
\def\eear          {\end{array}}
\def\eq            {\,{=}\,}
\newcommand\erf[2] {(\ref{#1#2})}
\def\ext           {^{\sss\rm ext}}
\newcommand\Fbox[1]{\fbox{$\,\displaystyle #1\,$}}
\def\findim        {fini\-te-dimen\-si\-o\-nal}
\def\fpc           {{{\cal F}_{\!\vec p,\tC}}}

\newcommand\Frac[2]{\mbox{\large$\frac{#1}{#2}$}}
\def\furu          {fusion rule}
\def\futnote#1     {\footnote{~#1}\ }

\def\g             {\mbox{$\liefont g$}}
\def\gb            {\mbox{$\bar{\liefont g}$}}
\def\G             {{\rm G}}
\def\Gc            {{\G_{\rm c}}}
\def\gpc           {{{\liefont g}^{}_{\vec p,\tC}}}

\def\Gs            {{G^*_{}}}
\def\Gz            {Generalized }
\newcommand\hsp[1] {\mbox{\hspace{#1 em}}}
\def\hy            {$\mbox{-\hspace{-.66 mm}-}$}
\def\I             {{\cal I}}
\def\ii            {{\rm i}}
\def\iN            {\,{\in}\,}
\def\infdim        {infinite-dimensional}
\def\Intro         {Introduction }

\def\irrep         {irreducible representation}
\def\J             {{\rm J}}
\def\JR            {{\rm J}}

\def\kz            {Knizh\-nik\hy Za\-mo\-lod\-chi\-kov}
\long\def\labl#1#2 {\label{#1#2}\ee}
\def\lambdab       {{\bar\lambda}}

\def\Lchir         {\mbox{{\small$\cal L$(\chir)}}}

\def\Lfloor        {\mbox{\normalsize$\lfloor$}}
\def\lie           {Lie algebra}
\def\Lie           {Lie group}
\def\liefont       {\mathfrak }
\def\llb           {\mbox{\large(}}
\def\lrb           {\mbox{\large)}}
\def\modinv        {modular invarian}
\def\mU            {{[\mu]}}
\def\Mu            {{[\mu,\psu]}}
\def\mub           {{\bar\mu}}

\def\muv           {{\vec\mu}}
\newcommand\N[3]   {{\rm N}_{#1,#2}^{\;\ \ #3}}
\def\nE            {\,{\ne}\,}

\newcommand\NN[3]  {{\rm N}_{#1,#2,#3}}
\newcommand\nnxt   {\raisebox{.12em}{\rule{.35em}{.35em}}\hsp{.6}}
\def\nup           {{\nu^{\!+}_{\phantom i}}}

\newcommand\nxt[1] {\\\nnxt{}#1}
\newcommand\Nxt[1] {\\[.25em]\nnxt{}#1}

\def\onedim        {one-dimen\-sional}

\def\ot            {\raisebox{.07em}{$\scriptstyle\otimes$}}

\def\otimeS        {\,{\otimes}\,}

\def\parfu         {partition function}
\def\Pe            {{\dl P}^1}
\newcommand\pho[1] {\phi^{}_{#1,\Tilde #1}}
\def\psu           {{\hat\psi}}
\def\Q             {Quantum }

\newcommand\rc[3]  {\CR^{#1}_{#2,\Tilde #2;#3}}
\newcommand\rca[3] {\CR^{#1_{\phantom i}}_{\! #2,\Tilde#2;#3^{\phantom|}_{}}}

\newcommand\rcaP[3]{\CR^{#1}_{\! #2,#2^+_{\phantom .};#3}}

\def\reals         {{\dl R}}
\def\rep           {representation}
\def\repth         {representation theory}
\def\resp          {respectively}
\def\Rfloor        {\mbox{\normalsize$\rfloor$}}
\def\rhob          {{\bar\rho}}

\def\sltwo         {\mbox{$\mathfrak{sl}(2)$}}

\def\SJ            {S^\J}

\def\slie          {Lie superalgebra}
\def\slz           {\mbox{SL(2,\Zet)}}
\def\sss           {\scriptscriptstyle}
\def\stt           {string theory}

\def\syms          {sym\-me\-tries}
\def\Tau           {\Theta}
\def\tC            {{\tilde C}}
\def\tft           {topological field theory}
\def\Tilde         {\dot}

\def\tmu           {{\tilde\mu}}

\def\To            {\,{\mapsto}\,}
\def\TO            {\mbox{$\Longrightarrow$}}

\newcommand\tNN[3] {\tilde{\rm N}_{#1,#2}^{\;\ \ #3}}

\newcommand\tNl[3] {\tilde{\rm N}_{#1,#2,#3}^{}}
\newcommand\tNL[3] {\tilde{\rm N}_{#1,#2,#3}}
\def\tS            {\tilde S}
\def\tvac          {{\tilde\vac}}
\def\twodim        {two-dimensional}
\def\vir           {\mbox{${\cal V}\!${\sl ir}}}
\def\vac           {\Omega}
\def\vacb          {{\bar\Omega}}

\def\voa           {vertex operator algebra}
\def\vop           {vertex operator}
\def\wfont         {\mathfrak }
\def\wrt           {with respect to }
\def\wrtt          {with respect to the }
\def\wzwm          {WZW model}

\def\Y             {\mbox{\sf Y}}
\def\zet           {{\dl Z}}
\def\Zet           {${\dl Z}$}

\def\zetplus       {\mbox{$\zet_{>0}$}}
\def\zetpluso      {{\zet_{\ge0}}}

\documentclass[12pt]{article} \usepackage{amssymb,amsfonts,latexsym}
\begin{document}
\begin{flushright}  {~} \\[-16mm] {\sf hep-th/0001005} \\[1mm]
{\sf PAR-LPTHE 99-48}\\[2mm] {\sf December 1999} \end{flushright}

\begin{center} \vskip 15mm
{\Large\bf BUNDLES OF CHIRAL BLOCKS AND\\[.53em] 
BOUNDARY CONDITIONS IN CFT $^{\rule{.25em}{.25em}}$}\\[17mm]
{\large\bf J\"urgen Fuchs}\\[2.5mm]
Institutionen f\"or ingenj\"orsvetenskap, fysik och matematik\\
Universitetsgatan 1\\ S\,--\,651\,88\,\, Karlstad\\[5mm] and\\[5mm]
{\large\bf Christoph Schweigert}\\[2.5mm] LPTHE, Universit\'e Paris VI\\
4 place Jussieu\\ F\,--\,75\,252\,\, Paris\, Cedex 05
\end{center}
\vskip 22mm
\begin{quote}{\bf Abstract}\\[1mm]
Various aspects of spaces of chiral blocks are discussed.
In particular, conjectures about the dimensions of irreducible
sub-bundles are reviewed and their relation to symmetry breaking
conformal boundary conditions is outlined.
\end{quote}
\vfill\noindent------------------\\
$^{\rule{.35em}{.35em}}$~Invited talk by J\"urgen Fuchs at the
Workshop on Lie Theory and Its Applications in Physics\\
\mbox{$\;\ $}\hy\ Lie III
(Clausthal, July 1999); slightly extended version of contribution
to the Proceedings.
\newpage


\section{Vertex operator algebras and chiral blocks}

Chiral blocks -- also known as {\em conformal\/} blocks -- arise in the
study of \twodim\ \cft. In physics terminology, they are
correlation ``functions" of so-called chiral vertex operators $\phi_\mu$.
Thus, roughly, one deals with objects of the form
  \be  \langle {\phi_{\mu_1}}{(p_1)}\, {\phi_{\mu_2}}{(p_2)}
  \cdots {\phi_{\mu_m}}{(p_m)} \rangle_{\!(\tC)} \,,  \labl01
where $\tC$ is a \twodim\ manifold, some kind of `operator product' between
the \cvo s $\phi_{\mu_i}$ `sitting' at $p_i\iN\tC$ is understood, and 
$\langle\cdots\rangle$ stands for
the operation of forming the `vacuum expectation value'.
Yet, \cb s are in general neither functions, nor uniquely 
determined by these data. For a more detailed understanding
several concepts are needed, among them in particular the following.
\Nxt
First, the notion of a {\em vertex operator algebra\/} 
$\chir\eq(\calhv{,}\Y{,}v_\vac{,}v_{\sss\rm Vir})$.
Here $\calhv\eq\bigoplus_n^{}\!\calhv^{(n)}$ is an \infdim\ \Zet-graded 
vector space, with \findim\
 \hepv{homogeneous subspaces }$\calhv^{(n)}$. 
$\calhv$ is endowed with infinitely many products, which are encoded 
in the {\em vertex operator map\/} $\Y{:}\ \calhv\,{\to}\,
{\rm End}(\calhv)\,{\otimes^{}_\complex}\,\complex[[t,t^{-1}]]$,
mapping $\calhv$ to the Laurent series in a formal variable $t$ with
values in the endomorphisms of $\calhv$. The {\em vacuum element\/} $v_\vac\iN
\calhv^{(0)}$ and the {\em Virasoro element\/} $v_{\sss\rm Vir}\iN\calhv^{(2)}$
are distinguished vectors in $\calhv$, satisfying
$\Y(v_\vac)\eq\mbox{\sl id}$ and $(\Y(v;t{=}0))v_\vac\eq v$
($\Y$ is therefore also known as state-field correspondence).
\\
These quantities are subject to a number of further
axioms\hepv{ (see e.g.\ \cite{FRlm,frhl,KAc4,scHl7})}, mostly not to
be spelled out here. We only mention the requirement that the endomorphisms 
$L_n$ defined by the expansion
$\Y(v_{\sss\rm Vir})\eq\sum_{n\in\zet}L_n\,t^{-n-2}$ form a basis of
the Virasoro algebra \vir. (More precisely, they provide a \rep\ of
\vir\ in which the central element acts as a constant multiple,
called the rank of \chir, of the identity.)
\nxt
For many purposes, it is sufficient to regard the vertex operator algebra as
the Lie \alg\ spanned over $\complex$ by the Fourier\hy Laurent modes of
suitable vertex operators $\Y(v)$. Besides \vir\ (i.e.\ the 
$L_n$) and its supersymmetric generalizations, examples are so-called
$\cal W$-algebras and untwisted affine \lie s\ \g. The term {\em chiral \alg\/}
is used both for the proper vertex operator algebra \chir\ and for the Lie \alg\ 
\Lchir.
\nxt
There is a collection of
irreducible \chir-modules $\calhm$, with $\mu$ in some index set $I$.
This includes $\vac\iN I$, i.e.\ the vector space $\calhv$ underlying \chir;
this is called the {\em vacuum sector\/}.
\\
The modules $\calhm$ are graded {\em weight\/} modules, with \findim\ weight 
spaces, where the weights are \wrtt zero mode $L_0$ of \vir\ and a
suitable collection $\{H^i_0\}$ of other mutually commuting modes in \Lchir.
Hence there is the notion of {\em characters\/}, i.e.\ generating functions 
$\chii_\mu(\tau,{\vec z})\eq{\rm tr}_{\calhm}
\eE^{2\pi\ii\tau L_0}\eE^{2\pi\ii{\vec z}{\cdot}{\vec H_0}}$
for weight multiplicities.
\nxt
When every \chir-module is fully reducible and $|I|\,{<}\,\infty$, one 
speaks of 
a {\em rational\/} \voa, \resp\ rational \Cft. Below we restrict to this case.
In a rational theory the modules $\calhm$ constitute the simple objects of a
modular tensor category (provided that
\chir\ is `maximal', which corresponds to non-degeneracy of braiding).

\medskip
We can now describe \cb s more properly; they are certain linear forms
  \be  B_\muv:\quad \calh_{\mu_1}\otimeS\calh_{\mu_2}\otimeS\cdots
  \otimeS\calh_{\mu_m}\equiv\, {\calhmv} \,\to \,{\complex}  \labl03
on the tensor product $\calhmv$ of the relevant irreducible modules
$\calh_{\mu_i}$. Further, to establish the connection with formula \erf01
one considers a complex curve $\tC$ with ordered marked points $p_i\iN\tC$,
and identifies for each $i$ the formal variable $t$ with 
a local holomorphic coordinate $\zeta_i$ at $p_i\iN\tC$. 
(For $\tC\eq\Pe$ one may take $t\;{\hat=}\;\zeta_i\eq z{-}z_i$ with
$z$ a quasi-global holomorphic 
coordinate on $\Pe$ -- say the standard global coordinate on $\complex$ 
for $\Pe\,{\sim}\,\complex\,{\cup}\{\infty\}$ -- such that $z(p_i)\eq z_i$. 
For higher genus curves, the prescription becomes more complicated.)
Then for each $v\iN\calhv$ the vertex operator map provides
a \cvo\ as appearing in \erf01, according to $\Y(v)\,\hat=\,\phi_\vac(v;
p_i)$. For other sectors $\mu\nE\vac$, the
\cvo s $\phi_\mu(v;p_i)$ with $v\iN\calhm$ correspond in an analogous 
manner to intertwining operators between \chir-modules.
Then one sets
  \be  \langle \phi_{\mu_1}(v_1{;}p_1)\, \phi_{\mu_2}(v_2{;}p_2)\,{\cdots}\,
  \phi_{\mu_m}(v_m{;}p_m) \rangle_{\!(\tC)}
  = B_\muv(v_1\ot v_2\ot\cdots\ot v_m) \,.  \labl04
Hereby the chiral \alg\ is interpreted as the ``local implementation of 
the symmetries" of the system at the {\em insertion points\/}
$\vec p\eq(p_1,p_2,...\,,p_m)$. But we also want to study \erf04 in its
dependence on the insertion points $\vec p$ and on the moduli
\hepv{$\vec\tau$ }of $\tC$. This necessitates the
construction, for each curve $\tC$ and number $m$ of insertions,
of a suitable ``{\em global\/} implementation of the symmetries".
Such an implementation, to be called a {\em block \alg\/}\,%
 \hepv{\futnote{Unfortunately, it is $\apc$ that mathematicians sometimes
 call the `chiral \alg', see e.g.\ \cite{gaiT}.}}
and denoted by $\apc$, is a family (varying with the 
 \hepv{insertion points and }moduli) of sub\alg s of 
the $m$-fold tensor product of \Lchir, and can be thought of \cite{fefk3}
as providing a generalized co-product.
 \hepv{

 }%
The action of $\apc$ on $\calhmv$ allows us to be specific about the
linear forms \erf03. Namely, one defines the chiral blocks to be the space
$B_\muv\eq\llb {(\calhmv)}^*_{\phantom I} \lrb^\apC_{\phantom I}$ of 
$\apc$-{\em singlets\/} in the algebraic dual ${(\calhmv)}^*$ -- or dually, 
as the space
  \be  \Fbox{ B_\muv^* = \Lfloor \calhmv \Rfloor_\apC^{\phantom I} }  \ee
of {\em co-invariants\/} of $\calhmv$ \wrt $\apc$.
In physics terminology, this prescription says that $B_\muv$ is the
space of solutions to the {\em Ward identities\/} of the system.

\section{\Cb s in \wzwm s}

By a {\em\wzwm\/} one means a \cft\ for which \Lchir\
is an untwisted affine \lie\ \g\ (or, more precisely, its
semi-direct sum with \vir) and for which the Virasoro \rep\
is supplied by the affine {\em Sugawara construction\/}, which says that
the Virasoro generators are quadratic expressions in the generators of \g,
with coefficients proportional to the Killing form of the horizontal
subalgebra $\g_0\,{\subset}\,\g$.

Various notions of \cft\ have very concrete WZW realizations:
\nxt
One can regard the affine \lie\ \g\ as being obtained via the {\em loop
construction\/} from a \findim\ simple Lie \alg\ \gb; the 
formal variable $t$ of the vertex operator formalism is closely related to
the indeterminate of the loop construction. In this presentation 
\g\ has a basis $\{J^a_n\}$ with $n\iN\zet$ and $\{J^a\}$ a basis of 
\gb\ (together with a central element $K$ and a derivation $D$), and
the simple \lie\ \gb\ can be identified with the horizontal subalgebra 
$\g_0$, which is spanned by the zero modes $J^a_0$.
\nxt 
The spaces $\calhm$ are irreducible highest weight modules over \g\
with integrable highest weight $\mu$ of fixed level $k\iN\zetplus$. 
There are only finitely many such weights $\mu$ (in fact the theory is 
rational), namely those whose horizontal part $\bar\mu$ is a dominant integral 
\gb-weight with inner product $(\bar\mu,\bar\theta^{\sss\vee}){\le}\,k$
with the highest coroot $\bar\theta^{\sss\vee}$. For instance, for
$\gb\eq\sltwo$ only the \sltwo-weights $\bar\mu\eq0,1,...\,,k$ are allowed.
 \\
The vacuum sector is the {\em basic\/} \g-module, which has highest weight
$\vac\equiv k\Lambda_{(0)}$.
\nxt
One has $L_0\eq{-}D$, and $\vec H_0$ form a basis of the Cartan sub\alg\ of
\gb. The characters $\chii_\mu$ are obtained from the corresponding formal 
\g-characters -- i.e.\ elements of the group \alg\ spanned by formal 
exponentials in the weights --
by interpreting the formal exponentials as functions on weight space.
They are convergent for $\Im(\tau)\,{>}\,0$.
\nxt
The block \alg\ $\apc$ is then (for details see e.g.\ 
\cite{beau,Ueno,scsh3}) the tensor product
  \be  \gpc = \gb \otimes \fpc \,,  \labl8u
with $\fpc$ the \alg\ of functions holomorphic on $\tC{\setminus}\{\vec p\}$ 
and with (at most) finite order poles at the $p_i$. The action of $\gpc$ 
on $\calhmv$ is given by
  $ \llb R_{\vec\mu}(\bar x\ot f)\lrb (v_1\ot v_2\ot\cdots\ot v_m)
  \,{:=}\, \sum_{i=1}^m v_1\ot v_2\ot\cdots\ot$
  $R_{\mu_i}(x_i) v_i\,\ot \cdots\ot v_m $
for $v_1\ot v_2\ot\cdots\ot v_m\iN\calhmv$, where the 
$x_i\,{\equiv}\,\bar x\ot f_{p_i}$ are to be regarded as elements 
of \g\ ($f_{p_i}$ denotes the local expansion of $f$ at $p_i$).

For general \Cfts, much less is known about
block \alg s and their action on tensor products. Roughly, one must `couple'
Virasoro-(quasi)primary fields to meromorphic sections of suitable
powers of the canonical bundle of $\tC$; in the WZW case this power is
zero, hence one deals with functions \erf8u and can be very explicit.
Thus for general \Cfts\ many facets of what is reported below
are not at all rigorous -- to us it is a major challenge in \Cft\
to improve this.\,%
 \futnote{It is as yet unclear whether the \vop\ framework
 is broad enough for a rigorous discussion of all issues of interest in
 \Cft, or whether one must resort to formulations involving e.g.\ von
 Neumann \alg s. (The latter would be unfortunate, as one would give up on
 treating non-unitary models, like ghost systems in string theory, at an
 equal footing as unitary ones.) In the 
 WZW case, such a formulation follows by studying the loop group ${\rm L}\G$
 of the compact, connected and simply connected real Lie group $\G$
 whose \lie\ is the compact real form of \gb, as well as the associated local
 loop groups and their \rep s on Hilbert spaces; compare e.g.\ \cite{xu3}.
 \\
 In this context, note that in the vertex operator setting no topology is
 chosen on the vector spaces $\calhl$, i.e.\ even in the unitary case they
 are only {\em pre\/}-Hilbert spaces. In fact, for certain purposes
 -- e.g.\ when trying to achieve that the generators of \Lchir\
 act continuously on the (dual) blocks -- other topologies than the 
 Hilbert space topology based on the standard norm can be more convenient.}
In contrast, for \wzwm s already enough is known so as to
make precise statements and establish rigorous proofs.


\section{Bundles of chiral blocks}

\Cb\ spaces have been studied in quite some detail for several reasons.
(In \Cft\ their significance emanates from the fact
that they contain the physical \corfu s as special elements, see below.)
Among the pertinent results are:
\nxt
In all known cases (not only for rational  \Cfts), for fixed insertion 
points $\vec p$ and fixed moduli $\vec\tau$ of $\tC$, the space
$B_\muv$ is a {\em\findim\/} vector space $B_\muv(\vec p,\vec\tau)$. 
This has a counterpart 
in the associated tensor category: all morphism spaces in a $C^*$-tensor 
category with conjugates and irreducible unit are \findim\ \cite{loro}.
\nxt 
The spaces $B_\muv(\vec p,\vec\tau)$ fit together to the total space 
of a finite rank vector bundle $\calb_\muv$ over the moduli space of 
genus $g$ complex curves with $m$ ordered marked points.
\nxt 
Using \vir\ one constructs a projectively flat
`\kz' connection on $\calb_\muv$. (Some authors reserve the 
term `chiral block' for flat sections of $\calb_\muv$.)
\nxt 
The block bundles are in general not irreducible
(i.e.\ the fibers decompose into a direct sum of
vector spaces, in a manner compatible with the transition functions). 
\nxt
One of the major reasons for independent mathematical interest in \cb s
is the role played in \alg ic geometry by the WZW one-point blocks with 
vacuum insertion $\phi_\vac$. Namely, the Picard group of 
the moduli space $\calm_{\G,\tC}$ of holomorphic principal
$\Gc$-bundles (with $\Gc$ the complexification of \G)
over $\tC$ modulo stable equivalence,\,%
 \futnote{$\calm_{\G,\tC}$ possesses several other interpretations as
 well, such as: the set of equivalence classes of flat principal \G-bundles;
 the space of semi-stable holomorphic vector bundles ${\rm E}$ over $\tC$ 
 such that the sheaf of sections of the determinant bundle is the structure 
 sheaf of $\tC$, ${\rm det}\,{\rm E}\eq{\cal O}_\tC$;
 and the phase space ${\cal A}_\circ/{\cal G}$ (flat connections modulo gauge
 transformations) of Chern\hy Si\-mons gauge theory.
 \\
 In the first place, these are just bijections of sets. But each 
 of the sets comes equipped with its own natural structures.
 One can translate those, so that indeed one gets a set with
 various different interesting structures. A crucial input for 
 establishing these relations is Borel\hy Weil\hy Bott theory.}
is generated by the determinant line bundle $\call$.
$\call\eq{\cal O}(\theta)$ is a locally free rank-one sheaf of meromorphic 
functions on $\calm_{\G,\tC}$, where $\theta$ is the Theta divisor.
Now for every $k\iN\zetplus$, the space of holomorphic
sections of $\call^{\otimes k}$ is canonically
isomorphic to the space of one-point blocks on $\tC$
with insertion $\phi_\vac$ at level $k$:
  \be  H^0(\calm_{\G,\tC},\call^{\otimes k}) \,\cong\, B_
  {k\Lambda_{(0)}}(\tC) \,.  \ee
With traditional methods this `{space of generalized Theta functions}'
had been accessible only in a few special cases
(for more information, see e.g.\ \cite{beau,SCho2}).

\section{Dimensions}

When studying \cb s, the first quantity of interest that comes to mind
is the rank of the bundle $\calb_\muv$, i.e.\ the dimension 
${\rm N}_{\muv;\tC}\eq{\rm dim}\,B_{\muv;\tC}$ of the spaces $B_{\muv;\tC}$.
In \Cft, the integers $\N\lambda\mu\nu\,{\equiv}\,\NN\lambda\mu{\nup;\Pe}$ 
give the {\em fusion rules\/}, i.e.\ the number 
$\#\,(\phi_\lambda{\star}\phi_\mu\,{\leadsto}\,\phi_\nu)$
of `independent couplings' between families of fields. Factorization
 (see section \ref{s.cf} below)
implies that
the fusion rules constitute the structure constants of a commutative
semi-simple associative \alg\ with unit and involution, which is called
the {\em fusion rule \alg\/}. They can be expressed in terms of a unitary
symmetric matrix $S$ by the {\em Verlinde formula\/}
  \be  \N\lambda\mu\nu = \sum_{\kappa\in I} {S_{\kappa,\lambda}^{}\,
  S_{\kappa,\mu}^{}\,S_{\kappa,\nu}^*} \,/\, {S_{\kappa,\vac}} \,.  \labl1v
It is worth pointing out that the existence of a 
\hepv{diagonalizing }matrix 
$S$ obeying \erf1v is an immediate by-product of the
representation theory of fusion rule \alg s. The contents of 
the {\em Verlinde conjecture\/} is not formula \erf1v in itself, 
but rather that it is one and the same matrix $S$ that appears in \erf1v and
that affords the modular trans\-for\-ma\-tion $\tau\,{\mapsto}\,{-}1/\tau$
on the characters $\chii_\mu$. This implies in particular
concrete expressions for $S$, e.g.\ the Kac\hy Peterson formula for
WZW models, and similarly for coset models and WZW orbifolds.
 \hepv{

}By factorization (see below), the Verlinde formula \erf1v generalizes as
  \be  \Fbox{ {\rm N}_{\muv;\tC} = \sum_{\kappa\in I} |S_{\kappa,\vac}|^{2-2g}
  \prod_{i=1}^m \frac{S_{\kappa,\mu_i}}{S_{\kappa,\vac}} }  \labl2v
to an arbitrary number $m$ of insertions and arbitrary genus $g$.
\erf2v has been proven rigorously only for \wzwm s (in particular by \alg ic 
geometry means, cf.\ e.g.\ \cite{beau,falt} and also \cite{tele,fink}).
But there is enormous evidence that it holds in general; in particular it
was verified for very many theories that the numbers \erf2v are in $\zetpluso$.

\section{Traces}\label{secC}

The dimensions \erf2v are only the most basic characteristics of 
blocks. Other quantities are, of course,
of interest as well. As the blocks are in general not irreducible as vector 
bundles, a natural generalization are the dimensions of irreducible sub-bundles.
For many \cb s, a non-trivial
sub-bundle structure follows from the presence of some group
$\cals$ of \auto s $\sigma$ of $\apc$, which in turn come from \auto s of \chir. 
There are then linear bijections $\Tau_{\sigma}$ between the $\calhm$
satisfying the twisted intertwiner property $\Tau_{\sigma}^{-1}\,{\circ}
\,\Y(\sigma v;z)\circ\Tau_{\sigma}^{}\eq\Y(v;z)$ and descending to 
linear maps $\Tau_{\vec\sigma}$ on the blocks. The 
$\Tau_{\vec\sigma}$ realize $\cals$ projectively, and the sub-bundles are
obtained by the simultaneous eigenspace decomposition of the blocks \wrt
these maps \cite{fuSc8}.

Some information on the dimensions of such sub-bundles is available, too.
The dimensions are most favorably expressed in terms of traces of the
twisted intertwiners $\Tau_{\vec\sigma}$, to which they are
related by Fourier transformation 
\wrtt subgroup of $\cals$ that corresponds to the center of a
twisted group algebra, where the twist is by the cocycle defining the 
projectivity of the action on the blocks. Concretely, there are 
generalizations of the Verlinde conjecture for two important types of \auto s: 
\nxt
First, for \auto s associated to
{\em simple currents\/} $\phi_\J$. A simple current is a unit of the fusion 
\alg; it can be characterized by the equality $S_{\J,\vac}\eq S_{\vac,\vac}$.
Simple currents of \wzwm s correspond to symmetries of the Dynkin
diagram of the underlying affine \lie\ \g\ and thereby to certain outer 
\auto s $\sigma_\J$ of \g. The proposed formula for the traces of the
 \hepv{corresponding }twisted intertwiners 
$\theta_{\vec{\J}}\,{\equiv}\,\theta_{\vec\sigma_\J}$ reads \cite{fuSc8}
  \be  \Fbox{
  {\rm Tr}_{\calb_{\muv;\Pe}}^{} \llb \Tau_{\J_1,\J_2,...,\J_m} \lrb
  = \sum_{\kappa:\ \J_\ell\star\kappa=\kappa}
  |{S_{\kappa,\vac}}|^2_{\phantom I}
  \prod_{i=1}^m \Frac{S_{\kappa,\mu_{\mbox{$\sss i$}}}
  ^{\J_{\mbox{$\sss i$}}}}{S_{\kappa,\vac}} \,. } \labl1C
Here it is assumed that $\J_1{\star}\J_2{\star}\cdots{\star}\J_m\eq\vac$
as well as $\J_i{\star}\mu_i\eq\mu_i$ for all $i\eq1,2,...\,,m$ (when
formally extended to other cases, the expression \erf1C just yields zero). 
Further, $S^\J$ is the modular matrix for one-point blocks 
with insertion $\phi_\J$ on an elliptic curve.
\nxt
Second, for \auto s $\sigma$ of \chir\ that preserve 
$v_{\sss\rm Vir}$ one finds \cite{fuSc12}
  \be  \Fbox{
  {\rm Tr}_{\calb_{\muv;\Pe}}^{} \llb \Tau_{\sigma,\sigma,...,\sigma} \lrb
  = \sum_{\dot\kappa} |S_{\dot\kappa,\vac}^\circ|^2_{\phantom i} 
  \prod_{i=1}^m \Frac{S_{\dot\kappa,\mu_{\mbox{$\sss i$}}}^\circ}
  {S_{\dot\kappa,\vac}^\circ} \,. } \labl2C
Here $S^\circ$ is an ingredient of the modular S-matrix for an {\em orbifold\/}
theory that is formed from the original CFT by quotienting out
$\sigma$. Note that $S^\circ$ has two distinct types of labels; they
correspond to the $\sigma$-twisted versus the untwisted sector of the orbifold.

In the WZW case, \erf1C is closely related to a Verlinde formula for
non-simply connected groups, see 
 \hepv{eq.\ }\erf so below,
while $S^\circ$ coincides \cite{bifs} with the ordinary S-matrix for (a pair of)
twisted affine \lie s (being `genuinely twisted' iff $\sigma$ is outer).

But both \erf1C and \erf2C are conjectured for arbitrary rational \Cfts\ --
they originate from structures present in every rational \Cft.
Even for WZW they are far from being proven rigorously -- a proof would in
particular imply a proof of the Verlinde formula itself. But there are definite
ideas for WZW models and also for some derived theories like coset models.
In addition, there is enormous numerical evidence: one obtains
non-negative integers for the ranks,\,%
 \futnote{A surprising empirical observation is that in the case of \erf1C
 the traces are actually
 integral themselves, even when the order of $\sigma_\J$ is larger
 than 2. This is reminiscent of a description of ${\rm dim}\,B_{\muv;\Pe}$
 as the Euler number of a suitable BGG-like complex, and hence 
 suggests that the traces may possess a homological interpretation as
 well. (On the other hand, the 
 acyclicity result that implies non-negativity of the dimensions cannot
 generalize. Similar structures have appeared in \cite{fema3}.)
}
even though they are obtained as complicated sums of arbitrary (well,
they all lie in a cyclotomic extension of $\dl Q$) complex numbers.

\section{Chiral versus full CFT}

The sub-bundle structure described by the result \erf2C can be understood
in the framework of orbifold \Cfts. As it turns out, the very same chiral 
concepts play a role in the study of symmetry breaking \bc s. Therefore
in the rest of this note we address this conceptually
different (and at first sight totally unrelated) topic. 

As a first step, let us point out that the whole
discussion so far concerns what we like to call
{\em chiral \cft\/}, that is,\,%
 \futnote{Sometimes the term `chiral \Cft' is used in a slightly
 different fashion.}
CFT on a compact \twodim\ manifold without boundary that has a {\em complex\/}
structure, or in short, CFT on a complex curve $\tC$.
The analytic properties of $\tC$ enter in particular in the definition 
of block \alg s. In contrast, in most applications in physics,\,%
 \futnote{Among them are string theory and many condensed matter phenomena. 
 But there do exist applications where it is
 chiral CFT proper that is relevant. An important example is three-\dim al
 topological field theory -- Chern\hy Simons theory in the WZW case --
 and thereby the (fractional) quantum Hall effect. 
 (While a priori in the quantum Hall effect there is thus no natural place
 for \modinv ce on the torus, arguments assigning a
 physical role to it were given in the literature.)}
one must consider CFT on a {\em real\/} \twodim\ manifold $C$\,%
 \futnote{You might have already wondered why the symbol
 $\tC$ was used above rather than just $C$.}
with {\em conformal\/} 
structure. $C$ may be non-orientable or have a boundary, and it
does not come with a natural orientation even when it is orientable.
We will refer to \Cft\ on $C$ as {\em full \cft\/}.
\\
While in chiral \Cft\ one deals with
the chiral \alg\ \chir, \cvo s, \cb s, characters, and fusion rules,
the key notions in full \Cft\ are
fields, \corfu s, the torus \parfu, operator products, and \bc s.

Chiral \Cft\ is of interest in its own right. But it also serves
as a convenient intermediate step in the analysis of full CFT,
since it allows to exploit the power of complex geometry.
At the geometrical level, the relation between chiral and full \cft\
is pretty simple. The surface $C$ possesses an oriented two-sheeted 
Schottky cover $\tC$, branched over the boundary $\partial C$,
from which one recovers $C$ by dividing out a suitable anticonformal 
involution $\I$. Here are the simplest examples:
\nxt
Orientable, no boundary: $C\eq S^2$ (sphere) \TO\ $\tC\eq\Pe{\sqcup}\Pe$,
$\,\I{:}\ (z,\Tilde z)\To(\Tilde z^*,z^*)$.
\nxt
Orientable, with boundary: $C\eq D^2$ (disk) \TO\ $\tC\eq\Pe$,
$\,\I{:}\ z\To1/z^*$.
\nxt
Non-orientable: $C\eq\dl{RP}^2$ (projective plane\,/\,`crosscap')
\TO\ $\tC\eq\Pe$, $\,\I{:}\ z\To{-}1/z^*$.

To implement the transition from $\tC$ to $C$ at the field theory level
requires more work; e.g.\ for connected $\tC$ the block \alg s are understood 
only in the simplest cases. For now suffice it to say that, in a rough sense,
in many respects the transition amounts to taking two copies of chiral objects.
In particular, each single (bulk) field on $C$ comes from
{\em two\/} \cvo s on $\tC$ (physically speaking, one has `image charges').
Thus it carries {\em two\/} chiral labels; we denote it by $\pho\mu$.
In addition one must impose some additional constraints and identifications,
to be given below.

Before proceeding, let us recall that for WZW models
various structures can be made fully explicit which for other classes
of \Cfts\ are not yet worked out in detail. Fortunately, this is an
issue mainly for chiral \Cft. Once the chiral theory is
taken for granted, considerations in full \Cft\ turn out to be
essentially model independent.

\section{Correlation functions} \label{s.cf}

\Corfu s are the `vacuum expectation values' of
suitable products of `(quantum) fields'. They constitute the quantities of 
most direct interest in applications. For instance, by integrating them
over moduli space one obtains string scattering amplitudes.
A message to be remembered is that fields and their \corfu s are
objects in {\em full\/} \Cft\ and hence `live' on the quotient 
$C$ of $\tC$, while chiral blocks `live' on $\tC$. Thus
the blocks can{\em not\/} be physical \corfu s; rather, a \corfu\ for
$C$ is a specific element in a corresponding space of blocks on $\tC$.\,%
 \futnote{Or what is the same, after choosing some (natural) basis in the block
 space: a specific linear combination of basis blocks. When $\tC$ is
 disconnected, this is usually written as a sesqui-linear combination
 of separate basis blocks for the two connected components of $\tC$.}
That element is determined by various constraints, coming in three types:
\nxt 
{}{\em Locality\/}: Correlators are (single-valued) functions 
of the insertion points $p_i$ -- unlike generic sections of the block
bundle, which typically is not a trivial vector bundle.
\nxt
{}{\em Locality}\,$'$: They are also functions of the moduli of $C$ 
(modulo the Weyl anomaly).
\nxt
{}{\em Factorization\/}: They are compatible with desingularization.
This amounts to a restriction on the allowed intermediate states
that contribute in singular limits, and is thereby closely related to the
existence of operator product expansions.
\\
Technically: $\tC$ is a stable 
algebraic curve with at worst ordinary double points as singularities. 
When $\hat C$ is a partial desingularization of $\tC$
that resolves a double point $p\iN\tC$ in two points $p',p''\iN\hat C$,
then factorization gives a canonical isomorphism 
  \be  \mbox{\Large$\bigoplus$}^{}_{\nu\in I}\,
  B_{\muv,\nu,\nup;\hat C} \,\cong\, B_{\muv,\tC} \,.  \ee
A priori neither existence nor uniqueness of a solution to
these constraints is clear.

A prominent example is provided by
the correlator for $m\eq0$ and $g\eq1$. Then $C$ is a
torus, and $\tC$ is the disconnected sum $E_\tau{\sqcup}E_{\Tilde\tau}$ of two 
elliptic curves with opposite orientation, $\Tilde\tau\eq{-}\tau^*$.
The 0-point correlator on $C$ is the torus \parfu\ $Z$, while a basis for
the 0-point blocks on $\tC$ are tensor products of (\vir-specialized) 
irreducible characters $\chii_\mu$ and $\Tilde{\chii}_{\Tilde\mu}$. So 
$Z$ is a sesqui-linear combination of characters:
  \be  Z(\tau) = \mbox{\Large$\sum$}^{}_{\mu,\Tilde\mu}\,
  Z_{\mu,\Tilde\mu}\, \chii_\mu(\tau)\, \llb\Tilde{\chii}_{\Tilde\mu}(\tau)
  \lrb^* \,.  \labl4z
$Z$ is highly constrained by \hepv{the property of }{\em modular invariance\/},
i.e.\ locality \wrtt modulus $\tau$ of $E_\tau$.
The solution of this constraint is of interest in its own right.

\section{Boundary conditions}

Solving the factorization and locality constraints is not easy at all, 
in general, and little is known about uniqueness.\,%
 \futnote{Just think of the case of the torus \parfu\ \erf4z, where the
 constraint (modular invariance) looks quite innocent, but is still hard to
 solve. See e.g.\ \cite{gann13} for review and references.}
But for special \corfu s, which are still of great interest,
a lot can be done explicitly and in much generality. An especially 
fortunate example is given by the
1-point functions $\langle\pho\mu\rangle$ for bulk fields on the disk.
These are important because, due to factorization, there is 
only a small number of basic building blocks. As long as only closed
orientable surfaces $C$ are studied, already the 3-point functions
on $S^2$ are sufficient. In the general situation,
in addition the 1-point functions $\langle\pho\mu\rangle$
on the disk and on $\dl{RP}^2$ are needed (as well as 3-point functions
for boundary fields, which correspond to open string \vop s) 
\cite{lewe3,fips,prss3,runk}.

The chiral blocks for $\langle\pho\mu\rangle$ are two-point blocks on the
Schottky cover $\Pe$ of the disk. Now one has
${\rm N}_{\mu,\Tilde\mu}\eq\delta_{\Tilde\mu,\mu^{\!+}_{\phantom i}}$, so 
only a single coefficient needs to be determined:
  \be  {{\langle\pho\mu}{(v\ot v')} {\rangle}}_{\!a}
  = {\rca a\mu\vac} {B_{\mu,\Tilde\mu}}{(v\ot v')} \,.  \ee
Again by factorization, the complex number $\rca a\mu\vac\!$ can be
interpreted as a {\em reflection coefficient\/}, which
appears in the bulk-boundary operator product
   \be  \pho\mu(z) \ {\sim}\ \mbox{\Large$\sum$}^{}_{\nu\in I}\,
   (1{-}|z|^2)^{-2\Delta_\mu+\Delta_\nu}_{}\, \rca a\mu\nu
   \Psi^{a,a}_\nu({\rm arg}\,z) \quad\ \mbox{for\,\ }|z|\,{\to}\,1 \,.  \labl16
For closed orientable $C$ it is generally expected
that the constraints possess a unique solution. In contrast,
the one-point functions on the disk are in general not unique,
but an additional label $a$ is needed.
This indicates that the disk can come with several distinct 
{\em boundary conditions\/} labelled by $a$.
A boundary condition is essentially the same as a consistent
collection of one-point functions of bulk fields on the disk.

One of the most fundamental tasks in CFT is to determine,
assuming the theory to be known at the chiral level,
all consistent conformally invariant \bc s.

\section{Classifying \alg s}

Thus let us address the task of determining the conformal \bc s for a \Cft\
that is known at the chiral level. Un\raisebox{.12em}{\tiny(?)}fortunately 
the requirement of conformal invariance is rather weak, simply because \chir\ 
is typically much larger than just \vir. As a result, there will
in general (e.g.\ already for free boson theories) be {\em infinitely many\/} 
conformal \bc s. Usually they will be difficult to survey.

A pragmatic way out of this dilemma is to impose invariance under all of 
\chir, or at least under a sufficiently large consistent chiral sub\alg\ 
$\chirb$ of \chir, rather than only under the Virasoro \alg\hepv{ \vir}. 
(Here `invariance' means that the behavior at the
boundary\hepv{ $\partial C$}, cf.\ formula \erf16, is
identical for all bulk fields that are associated to vectors in a given
$\chirb$-submodule of an \chir-module $\calhm$.) Then
one can achieve a `rational' situation, with only finitely many \bc s.
Now by comparing two different factorization limits of the two-point function
${\langle\pho\mu\,\pho\nu\rangle}_{\!a}$, one can show that
 \be  \rca a\lambda\vac \rca a\mu\vac
  = \mbox{\Large$\sum$}^{}_\nu\, \tNN\lambda\mu{\,\nu}\, \rca a\nu\vac \ee
with numbers $\tNN\lambda\mu\nu$ which are combinations of fusing matrices 
and operator product coefficients. At first glance, these 
expressions look very complicated. But there is a crucial insight:
manifestly, $\tNN\lambda\mu\nu$ does not dependent on the boundary 
condition $a$.

This observation allows us to interpret
the reflection coefficients {$\rc a\mu\vac$} as furnishing
a \onedim\ \irrep\ of an \alg\ $\cal C$(\chirb) with structure constants 
$\tNN\lambda\mu\nu$, termed \cite{fuSc6} the {\em \cla\/}.
The results of \cite{card9} may be summarized by the statement
that the \cla\ \clA\ for boundary conditions preserving the full bulk 
symmetry \chir\ (and with charge conjugation as torus \parfu)
is nothing but the fusion \alg\ of the CFT. Thus 
\clA\ is a semi-simple associative \alg, its
structure constants are expressible through the Verlinde matrix $S$ as 
in \erf1v, and both a basis of \clA\ and the boundary conditions $a$ 
are labelled by the set $I$ of chiral labels $\mu$. 
(Yet, an explicit verification of $\tNN\lambda\mu\nu\eq\N\lambda\mu\nu$
was achieved \cite{prss3} only in special cases
where the relevant operator products and fusing matrices are known.)
  \hepv{
 
  }%
When $\chirb\ne\chir$, \hepv{then }the situation is more complicated, though
the factorization arguments go through.
For such {\em symmetry breaking \bc s\/} one finds:
\nxt 
One still has \onedim\ \irrep s of some \alg\ $\calc\eq\calc(\chirb)$.
\nxt
But the \corfu s are different. Namely, they are formed as different
combinations of the \cb s for \chir-descendant fields that are \chirb-primaries.
\nxt 
The labelling $\{\tilde\mu\}$ of basis elements of $\cal C$ 
and $\{a\}$ of \bc s is more subtle. In particular the two sets of labels 
are distinct; both differ from the set $I$.
\nxt
When
the unbroken part of the bulk symmetries constitutes the fixed point \alg\
  \be  \chirb_{\phantom |} = \chir^G  \labl18
\wrt any finite abelian group $G$ of automorphisms of \chir, then the
\bc s can be analyzed via $G$-orbifold and simple current techniques.

\hepv{\newpage}
\section{Interlude: Simple current extensions}\label{secJ}

One of the CFT concepts that was
instrumental for arriving at conjectures \erf1C and \erf2C is the
{\em simple current extension\/} of a rational CFT. It will show up again in the
study of \bc s below. Assume that the following data are given:\,%
 \futnote{While usually this is formulated by saying that one has some \Cft\
 with corresponding properties (and indeed there {\em are\/} many \Cfts\ with
 those properties), here we need not refer directly to CFT.}
\nxt
A set $\{ {\chii_\mu} \}$ \,($\mu\iN I$, $|I|\,{<}\,\infty$)\,
of functions of $\tau\iN\complex$, convergent for $\Im(\tau)\,{>}\,0$ and
forming a basis of a unitary module $V$ over \slz\ for which
$S\eq S^{\rm t}$ and $T\eq{\rm diag}$.
\nxt 
A vacuum label $\,\vac\iN I\,$, satisfying $\,S_{\vac,\mu}\iN\reals_{>0}\,$
for all $\mu\iN I$, and an involution $\,\mu\,{\mapsto}\,\mu^+\,$ on $I$ such that
$\,\vac^+\eq\vac\,$ as well as $S_{\lambda,\mu^+}\eq S^*_{\lambda,\mu}$ and 
$T_{\mu^+}\eq T_\mu$ for all $\lambda,\mu\iN I$.
\nxt 
A subset $\calg\,{\subseteq}\,I\,$ such that
$S_{\J,\vac}\eq S_{\vac,\vac}$ and $T_\J\eq T_\vac$ for all $\J\iN\calg$.
\\
(In CFT terms: $\J\iN\calg$ has the same quantum dimension (namely unity) and
the same conformal weight $\bmod\,\zet$ (namely zero) as $\vac$, i.e.\ is 
an integer spin simple current.)
\nxt
The numbers $\N\lambda\mu\nu$, regarded as {\em defined\/} by formula \erf1v,
are non-negative integers. 

In this situation one defines a {\em fusion ring\/} with product `$\star$'
on the vector space spanned by $\{\varphi_\mu\,|\,\mu\iN I\}$
by $\varphi_\lambda\,{\star\,}\varphi_\mu\,{:=}\,\sum_{\nu\in I}
\NN\lambda\mu{\nu^+}\,\varphi_\nu$, and can show rigorously \cite{scya6}:
\nxt 
$\calg$ is a finite abelian group w.r.t.\ `$\star$'
-- the group of units of the fusion ring.
\nxt 
$\calg$ organizes $I$ into orbits $[\mu]:=\{ \J\mu\,|\,\J{\iN}\calg \}$, with\,
$\phi_\J\star \phi_\mu \,{=:}\, \phi_{\J\star\mu} \,{\equiv}\, \phi_{\J\mu}$.
\nxt 
Defining the {\em stabilizer subgroup\/}
$\cals_\lambda\,{:=}\,\{\J\iN\calg\,|\,\J\lambda\eq\lambda\}$, the combination
  \be  Z{(\tau)}
  = \sum_{{\mU}: \ {\mu\in I},\ \hepv{ \quad \atop}
  {T_{\J\mu}=T_\mu}\ \forall\J\in\calg}\!\!
  \llb\, |{\cals_\mu}|\cdot \mbox{\Large$|$}\! \sum_{{\J\in\calg/\cals_\mu}}\!
  {\chii_{\J\mu}}{(\tau)} \mbox{\Large$|$}^2 \,\lrb  \labl21
is \slz-invariant. $Z$ is called a simple current extension 
modular invariant.\,%
 \futnote{Many of these modular invariants are interesting. Examples include 
the $D_{\rm even}$ type invariants of the \sltwo\ \wzwm\ and the invariant
 $Z\eq |{\chii_1 \,{+}\, \chii_{35}\,{+}\,\chii_{35'}\,{+}\,\chii_{35''} }|^2
 + 4\, |{\chii_{28}}|^2 $ for $D_4$ level 2.  \label{f10}}

To justify this name, one must be able to interpret \erf21 as the diagonal
invariant for some extended \Cft. This was achieved in \cite{fusS6}, where the
following was proven:
\nxt 
The extended labels are equivalence classes of pairs $\Mu$ with 
$T_{\J\mu}\eq T_\mu$ and $\psu$ a character of the {\em untwisted stabilizer\/}
${\calU_\mu}\,{:=}\,\{\J\iN\cals_\mu \,|\, {F_\mu(\J{,}\J'){=}1}
\,\forall\, {\J'}{\in}\cals_\mu\}\,{\subseteq}\,\cals_\mu$. Here $F_\mu$ 
is an alternating bihomomorphism on $\cals_\mu$, and hence 
 \hepv{the commutator cocycle }$F_\mu(\J{,}\J')\eq{\cal F}_\mu
 \linebreak[0](\J{,}\J')/{\cal F}_\mu (\J'{,}\J)$
for some\hepv{ cohomology class} ${\cal F}_\mu\iN H^2(\cals_\mu{,}{\rm U}(1))$
\cite{fusS6,bant7,muge6,fuSc11}. Thus the group \alg\ $\complex\,\calU_\mu$ 
is isomorphic to the center of the twisted group \alg\ 
$\complex_{{\cal F}_\mu}\!\cals_\mu$, implying that the inclusion 
$\calU_\mu\,{\subset}\,\cals_\mu$ is of square index $d_\mu^2$, with 
$d_\mu$ the dimension of the irreducible
$\,\complex_{{\cal F}_\mu}\!\cals_\mu$-\rep s.\,%
 \futnote{$F_\mu$ enters in calculations at various places; 
 that for $F_\mu\,{\not\equiv}\,1$ everything still nicely fits together
 \cite{fuSc11,fuSc12} is a strong consistency check. Non-trivial $F_\mu$
 appear naturally via products of simple currents that individually have 
 $T_\J\eq {-}T_\vac$, e.g.\ when one deals with tensor products of subtheories,
 such as in Gepner type string compactifications.
 \hepv{ \\ }In
 the $D_4$ example of footnote \ref{f10}, one finds
 $\cals_{28}\eq\calg\eq\zet_2{\times}\zet_2$ but
 $\calU_{28}\eq\{\vac\}$. Thus for $\mu\eq\Lambda_{(2)}\,{\hat=}\,28$ there is
 only a single extended character $\chii\ext_{28}\eq2\,{\cdot}\,\chii_{28}$.}
\nxt 
The summands in \erf21 are to be
read as $|\calU_\mu|\,{\cdot}\,|\chii\ext_\mU|^2$, i.e.\ for each
$[\mu]$ there are $|\calU_\mu|$ many extended irreducible characters
$\chii\ext_{[\mu,\psu]}$. Correspondingly the decompositions 
  \be  \calh\ext_{[\mu,\psu]} = \mbox{\Large$\bigoplus$}^{}
  _{\J\in\calg/\cals_\mu} \complex^{d_\mu} {\otimes} \calh_{\J\mu}  \ee
hold, and $\chii\ext_{[\mu,\psu]} \eq d_\mu{\cdot} \sum_{\J\in\calg/
\cals_\mu} \chii_{\J\mu}$ is the character of the extended module
$\calh\ext_{[\mu,\psu]}$.
\nxt 
Given, for every $\J\iN\calg$, a 
unitary matrix $\SJ{}$ satisfying the \slz\ relations
as well as $S^\J_{\lambda,\mu}\eq S^{\J^{-1}}_{\mu,\lambda}$
($\J\iN\cals_\lambda{\cap}\cals_\mu$) and $S^\vac\eq S$, the modular 
S-transformation matrix $S\ext$ of the functions $\chii\ext_{[\mu,\psu]}$
is obtained by sandwiching the $S^\J$ between group characters:
  \be  \Fbox{
  S\ext_{[\lambda,\psu_\lambda],[\mu,\psu_\mu]}
  = \,\Frac{|\calg|^{\phantom|}}{[\,|\cals_\lambda|\,|\calU_\lambda|\,
  |\cals_\mu|\,|\calU_\mu|\,{]}^{1/2}_{\phantom I}}
  \sum_{\J\in\calU_\lambda\cap\calU_\mu}
  \psu_\lambda(\J)\, \SJ_{\lambda,\mu}\, {\psu_\mu}(\J)^* \,. } \labl so
One has $\SJ_{\J'\lambda,\mu}\eq(T_\mu/T_{{\J'}\mu})\,{F_\mu}({\J},{\J'})
\,\SJ_{\lambda,\mu}$.
$S\ext$ is proven to be unitary and symmetric and to satisfy the \slz\
relations, and it was checked in a huge number of examples that it produces 
non-negative integers when inserted in the Verlinde formula \cite{fusS6}.
\nxt 
There is evidence \cite{bant7} that $\SJ$ is the modular S-matrix 
for the one-point blocks on the torus with insertion $\J$.
For WZW or coset models, $\SJ$ is the
Kac\hy Peterson matrix of the {\em orbit \lie\/} that is related\,%
 \futnote{See \cite{fusS3,furs}, and also
 \cite{muhl,bant7,schW3,brho,bofm,wend2,nait11,kakk}
 for background material and related work.}
to \g\ by a folding of the Dynkin diagram.
\nxt
Simple currents of \wzwm s correspond to the elements of the center of the
relevant covering group \G. It follows that \erf so appears in the
Verlinde formula for non-simply connected groups. (This result
was checked in \cite{beau3} for some simple cases.)

\section{The \cla\ for finite abelian $G$}

A systematic classification of \bc s has been achieved for all cases where
\chirb\ is given as in \erf18, with finite abelian automorphism group $G$ 
\cite{fuSc11,fuSc12}. (The simplest case $G\eq\zet_2$ includes e.g.\
Dirichlet \bc s for free bosons.) 

All basic ingredients are already known\,%
 \futnote{In particular one can make use of the fact that
 simple current extension by $\calg\cong\Gs$ provides the inverse
 operation to forming the orbifold \wrtt finite abelian group $G$.
 This way one can exploit both orbifold techniques and
 the simple current framework sketched in section \ref{secJ}.}
from chiral CFT. In particular:
\nxt
The label sets $\{\tmu\}$ for the basis of $\calc(\chirb)$ and $\{a\}$
for \bc s arise as two different deviations from the labels appearing in
\erf so. For $\tmu$, one has a character of
$\cals_\mu$ rather than of $\calU_\mu$, and no orbit is to
be taken, but still the requirement $T_{\J\mu}\eq T_\mu$ is kept, while
$a$ has the same form as extended labels, but now $T_{\J\mu}\,{\ne}\,T_\mu$
is allowed:
  \be  \Fbox{ \tmu = (\mub,\psi) } \qquad {\rm and} \qquad
  \Fbox{a = [\rhob(a),\psu{]}^{\phantom i}_{\phantom i} }  
  \qquad {\rm with}\;\ \psi\iN\cals_\mub^*,\;\psu\iN\calU_{\rhob(a)}^*  \labl56
(recall $\calU_\lambdab\,{\subseteq}\,\cals_\lambdab\,{\subseteq}\,\calg$).
\erf56 follows by heuristic considerations resembling ideas in \cite{card9}.
\nxt
Comparing with \erf so, we can
make an educated guess for a diagonalizing matrix:
  \be  \Fbox{
  \tS_{(\lambdab,\psi_\lambda),[\rhob,\psu_\rho]}
  = \Frac{|\calg|^{\phantom|}_{}} {[\,|\cals_\lambdab|\,|\calU_\lambdab|\,
  |\cals_\rhob|\,|\calU_\rhob|\,{]}^{1/2}_{\phantom I}}
  \sum_{\J\in\cals_\lambdab\cap\calU_\rhob}
  \psi^{}_\lambdab(\J)\, \SJ_{\lambdab,\rhob}\, \psu^{}_\rhob(\JR)^*\,. }
  \labl tS
\nnxt
Then \clAb\ is {\em defined\/} by prescribing the Verlinde-like formula
featuring $\tS$:
  \be  \tNl{\tilde\lambda}{\tilde\mu}{\tilde\nu}
  := \mbox{\Large$\sum$}^{}_a\, \tS_{\tilde\lambda,a} \tS_{\tilde\mu,a}
  \tS_{\tilde\nu,a} \,/\, {\tS_{\tilde\vac,a}} \,.  \labl tN
The structure constants are obtained from \erf tN by
raising the third index via $\tNL{\tilde\lambda}{\tilde\mu}{\tilde\vac}$.

\smallskip
There is as yet no rigorous derivation of formula \erf tS. But once
\erf tS and \erf tN are taken for granted, \clAb\ can be studied with
full rigor. In particular one shows \cite{fuSc11,fuSc12}:
\nxt
$\tS$ is (weighted) unitary. (Note that it is even non-trivial that 
$\tS$ is a square matrix.)
\nxt 
\clAb\ is a semi-simple commutative associative \alg\ with unit element 
$\tilde\vac\eq\vacb$\hepv{ (the vacuum sector of the $G$-orbifold)}.
\nxt 
The structure constants $\tNN{\tilde\lambda}{\tilde\mu}{\;\tilde\nu}$ of
\clAb\ are diagonalized by the matrix \erf tS. The irreducible
\clAb-\rep s $R_a$ are \onedim\ and labelled by the boundary 
labels $a$; they yield the reflection coefficients as
$\rcaP a\tmu\tvac\eq R_a(\phi_\tmu)\eq\tS_{\tmu,a}/\tS_{\tvac,a}$.
\nxt 
As an \alg\ over $\complex$, \clAb\ decomposes into ideals as
  \be  {\cal C}(\chirb)\,\cong\,\mbox{\Large$\bigoplus$}^{}_{\!g\in G}\;
  {\cal C}^{(g)}(\chirb) \,.  \labl cg
The ideal ${\cal C}^{(g)}(\chirb)$ plays the role of a \cla\ for
\bc s of definite {\em automorphism type\/} $g$. The 
corresponding boundary states are linear combinations of
$g$-twisted \cb s, which obey $g$-twisted Ward identities.\,%
 \futnote{For consistent subalgebras that are not fixed point \alg s,
 there exist \bc s which do not possess an automorphism type. Examples of such
 \bc s are e.g.\ known for the $\zet_2$-orbifold of a free boson and for
 the $E_6$-type invariant of the \sltwo\ \wzwm.}
\nxt
The ideal ${\cal C}^{(e)}(\chirb)$ appearing in \erf cg
is precisely the fusion rule \alg\ of \chir.
\nxt
It is plausible that orbifolding 
can be understood in terms of the folding of fusion graphs, and that
\hepv{the classifying \alg\ }${\cal C}(\chirb)$
thus coincides with the corresponding Pasquier \alg\ \cite{bppz2}.\,%
 \futnote{At least for cyclic $G$ -- for non-cyclic $G$ one must be aware of
 the possibility of having non-trivial 
 \hepv{two-cocycles }${\cal F}_\mu$.
 \hepv{ \\ }%
 Also, in practice this is difficult to check, because the Pasquier \alg\ 
 is obtained by an algorithm which does not directly produce uniform formulae
 for all rational \Cfts. The identification seems to be established
 so far only for $G\eq\zet_2$ in \sltwo\ WZW models and Virasoro minimal models.}
\nxt
The statements above refer to the charge conjugation torus \parfu. More
general results follow via 
{\em T-duality\/} symmetries, \hepv{which are }similar to those of free boson
theories, acting compatibly on the \bc s and on the torus \parfu.

\smallskip
Knowing 
 \hepv{the \cla\ }\clAb\ 
explicitly, a variety of consistency checks can be made.
Most importantly, one can prove \cite{fuSc11} integrality of the
coefficients of characters in the annulus amplitude (open string \parfu). 
This integrality is often used as the starting point
for studying boundaries; here it rather serves as an independent check.

Finally we remark that one can express the structure constants 
$\tNN{\tilde\lambda}{\tilde\mu}{\;\tilde\nu}$ as well as the annulus coefficients
through traces of twisted intertwining operators $\Tau_{\vec\sigma}$ on \cb\
spaces.  This yields the announced connection to the topic studied in section 
\ref{secC}.

\section{Conclusions and outlook}

Let us summarize by telling what we regard as the two main messages:
\nxt
First, there exists a close relation between the sub-bundle structure of 
chiral blocks (leading to the trace formula \erf2C) and symmetry 
breaking \bc s.
\nxt
Second, there is a systematic \class\ of all \bc s leaving unbroken a 
fixed point \alg\ $\chirb\eq\chir^G$
\wrt an arbitrary finite abelian group $G$. Concretely, one has
a general prescription, valid for all rational \Cfts, for the classifying 
\alg, with structure constants expressed through known {\em chiral\/} data.
\\[.3em]
Our results illustrate that the
space of \bc s has a rich and unexpectedly nice structure. We 
believe that many more issues are accessible quantitatively.
\vskip.3em
 
\noindent
Among possible extensions of the work outlined above we mention:
\nxt
One should find the diagonalizing matrix $\tS$ of \clAb\
when $\chirb\,{\ne}\,\chir^G$ for any group $G$.
\nxt
2-d \bc s can be understood in terms of 3-d topological theory \cite{fffs2}.
\nxt
Non-orientable surfaces, e.g.\ one-point functions on 
$\dl{RP}^2$ and the \parfu s of the Klein bottle and the M\"obius strip, 
are studied in \cite{prss3} and \cite{huss,fffs2,gann14}.
\nxt
One should look for a geometric interpretation of \bc s for 
non-flat backgrounds. For \wzwm s this is indeed available: one obtains
`fuzzy' versions of (possibly twisted) conjugacy classes of the group 
manifold \G\ \cite{alsc2,alrs3,Gawe,fffs,staN5}.
\nxt
More \hepv{explicit }information on the chiral data for further classes of 
models is highly welcome.
\nxt
Applications to string theory include, e.g., the complete analysis of
concrete compactifications
and a more systematic understanding of tadpole cancellation.


 \newcommand\wb{\,\linebreak[0]} \def\wB {$\,$\wb}
 \newcommand\Bi[1]    {\bibitem{#1}}
 \renewcommand\J[5]   {{\sl #5}, {#1} {#2} ({#3}) {#4}}
 \newcommand\Prep[2]  {{\sl #2}, preprint {#1}}
 \newcommand\PRep[2]  {{\sl #2}, preprint {#1}}
 \newcommand\BOOK[4]  {{\em #1\/} ({#2}, {#3} {#4})}
 \newcommand\inBO[7]  {{\sl #7}, in:\ {\em #1}, {#2}\ ({#3}, {#4} {#5}), p.\ {#6}}
 \def\jf    {J.\ Fuchs}
\hepv{\def\AMS    {{American Mathematical Society}}}
 \def\AP     {{Academic Press}}
 \def\Be     {{Berlin}}
 \def\Ca     {{Cambridge}}
 \def\CUP    {{Cambridge University Press}}
 \def\MD     {{Marcel Dekker}}
 \def\NY     {{New York}}
 \def\PR     {{Providence}}
 \def\Si     {{Singapore}}
 \def\SV     {{Sprin\-ger Ver\-lag}}
 \def\WS     {{World Scientific}}
 \def\atmp  {Adv.\wb Theor.\wb Math.\wb Phys.}
 \def\coma  {Con\-temp.\wb Math.}
 \def\comp  {Com\-mun.\wb Math.\wb Phys.}
 \def\gafa  {Geom.\wB and\wB Funct.\wb Anal.}
 \def\ijmp  {Int.\wb J.\wb Mod.\wb Phys.\ A}
 \def\inma  {Invent.\wb math.}
 \def\jgap  {J.\wb Geom.\wB and\wB Phys.}
 \def\jhep  {J.\wb High\wB Energy\wB Phys.}
 \def\joag  {J.\wB Al\-ge\-bra\-ic\wB Geom.}
 \def\joal  {J.\wB Al\-ge\-bra}
 \def\lmslns{London Math.\ Soc.\ Lecture Note Series \# }
 \def\mams  {Mem.\wb Amer.\wb Math.\wb Soc.}
\hepv{\def\mams  {Memoirs\wB Amer.\wb Math.\wb Soc.}}
 \def\nupb  {Nucl.\wb Phys.\ B}
 \def\phlb  {Phys.\wb Lett.\ B}
 \def\phrd  {Phys.\wb Rev.\ D}
 \def\phrl  {Phys.\wb Rev.\wb Lett.}
 \def\slnp  {Sprin\-ger\wB Lecture\wB Notes\wB in\wB Physics}
 \newcommand\geap[2] {\inBO{Physics and Geometry} {J.E.\ Andersen, H.\
            Pedersen, and A.\ Swann, eds.} \MD\NY{1997} {{#1}}{{#2}} }
 \newcommand\mbop[2] {\inBO{The Mathematical Beauty of Physics}
            {J.M.\ Drouffe and J.-B.\ Zuber, eds.} \WS\Si{1997} {{#1}}{{#2}} }
 \def\A     {Algebra}

\hepv{
\def\con           {conformal }
\def\kma           {Kac\hy Moody algebra}
\def\kna           {Krichever\hy Novikov algebra}
\def\kze           {Knizh\-nik\hy Za\-mo\-lod\-chi\-kov equation}
\def\mimo          {minimal model}
\def\Rep           {Representation}
\def\WZ            {Wess\hy Zu\-mino }
}
\hepv{\newpage  \small }

\end{document}